\documentclass[a4paper,fleqn,usenatbib]{mnras}

\usepackage{mathptmx}

\usepackage[T1]{fontenc}
\usepackage{ae,aecompl}

\usepackage{graphicx}	
\usepackage{amsmath}	
\usepackage{amssymb}	
\usepackage{setspace}

\newcommand{\pcm}{cm$^{-2}$}
\newcommand{\xmm}{\textit{XMM-Newton}}

\newcommand{\delcstat}{$\Delta$C-stat}

\title[Ultra-fast outflow in PG 1448+273]{Detection of a variable ultra-fast outflow in the Narrow Line Seyfert 1 galaxy PG 1448+273}

\author[P Kosec et al.]{P. Kosec$^{1}$\thanks{E-mail: pk394@cam.ac.uk}, 
A. Zoghbi$^{2}$,
D. J. Walton$^{1}$,
C. Pinto$^{3}$,
A. C. Fabian$^{1}$,
M. L. Parker$^{4}$ \newauthor
and C. S. Reynolds$^{1}$
\\
$^{1}$Institute of Astronomy, Madingley Road, CB3 0HA Cambridge, UK \\
$^{2}$Department of Astronomy, University of Michigan, Ann Arbor, MI 48109, USA\\
$^{3}$INAF – IASF Palermo, Via U. La Malfa 153, I-90146 Palermo, Italy\\
$^{4}$European Space Astronomy Centre (ESAC), Science Operations Department, 28692 Villanueva de la Ca\~nada, Madrid, Spain\\
}

\date{Accepted XXX. Received YYY; in original form ZZZ}

\pubyear{2020}

\begin{document}
\label{firstpage}
\pagerange{\pageref{firstpage}--\pageref{lastpage}}
\maketitle

\begin{abstract}

Relativistically blueshifted absorption features of highly ionised ions, the so-called ultra-fast outflows (UFOs), have been detected in the X-ray spectra of a number of accreting supermassive black holes. If these features truly originate from accretion disc winds accelerated to more than 10 per cent of the speed of light, their energy budget is very significant and they can contribute to or even drive galaxy-scale feedback from active galactic nuclei (AGN). However, the UFO spectral features are often weak due to high ionisation of the outflowing material, and the inference of the wind physical properties can be complicated by other spectral features in AGN such as relativistic reflection. Here we study a highly accreting Narrow Line Seyfert 1 galaxy PG 1448+273. We apply an automated, systematic routine for detecting outflows in accreting systems and achieve an unambiguous detection of a UFO in this AGN. The UFO absorption is observed in both soft and hard X-ray bands with the \xmm\ observatory. The velocity of the outflow is $(26900 \pm 600)$ km/s ($\sim$0.09c), with an ionisation parameter of $\log (\xi / \textrm{erg cm s}^{-1})=4.03_{-0.08}^{+0.10}$ and a column density above $10^{23}$ \pcm. At the same time, we detect weak warm absorption features in the spectrum of the object. Our systematic outflow search suggests the presence of further multi-phase wind structure, but we cannot claim a significant detection considering the present data quality. The UFO is not detected in a second, shorter observation with \xmm, indicating variability in time, observed also in other similar AGN.

\end{abstract}




\section{Introduction}

Highly blueshifted absorption lines of ionised material have recently been detected in the X-ray spectra of a number of active galactic nuclei (AGN), reaching velocities from $\sim$0.1c to as high as 0.5c \citep[e.g.][]{Pounds+03, Reeves+03, Tombesi+10b, Tombesi+10a, Tombesi+15, Nardini+15, Parker+16, Kosec+18c, Walton+19}. These absorption lines could originate from accretion disc winds, launched at high velocities from the accretion flow of AGN by radiation pressure \citep{Proga+00} or magnetic forces \citep{Fukumura+10}. The large velocity and the consequent significant kinetic power makes these outflows strong candidates for being the drivers of AGN feedback in galaxies \citep{Fabian+12, King+15}. Alternatively, these blueshifted absorption lines could originate from low-density material co-rotating with the inner accretion flow without the need for an outflow from the system \citep{Gallo+13, Fabian+20a}.

In most cases the signatures of these so-called `ultra-fast outflows' (UFOs) were identified using the highly ionised features of \ion{Fe}{XXV} and \ion{Fe}{XXVI} \citep{Tombesi+10b} in the iron K band (7-10 keV). The iron K band, however, is often on the upper edge of the energy range of the current instruments like XMM-Newton (EPIC pn and MOS) and Chandra, where the collecting area of the instruments steeply drops off, resulting in poor counts and signal-to-noise ratios. Additionally, this band commonly contains strong features of relativistically blurred reflection of the AGN coronal emission from the accretion disc \citep{Tanaka+95, Fabian+09}. Therefore the observed absorption features can be hard to interpret as outflow signatures and it is difficult to infer the wind properties such as the velocity and the ionisation level \citep{Zoghbi+15}. Furthermore, spurious low significance features or lines of instrumental origin can be interpreted as real UFO signatures.

Less frequently, fast ionised wind absorption lines are also observed in softer X-rays ($<5$ keV), for example using the \ion{Si}{XIV}, \ion{S}{XVI} and \ion{O}{VIII} transitions \citep{Jiang+18, Pinto+18, Reeves+18a}. The soft X-ray band (0.3-2.0 keV) can however be dominated by the commonly observed warm absorption from low ionisation material at larger distances from the AGN \citep[e.g.][]{Kaastra+00,Kaspi+02,Blustin+05,Detmers+11}, which complicates line identification. 

It is therefore important to study UFOs in different objects and across a broad X-ray band, comparing their properties to the source X-ray continuum properties in order to understand the true nature of UFOs and their relationship to AGN.

Here we present the detection of an ultrafast outflow with a velocity of 0.09c in the spectrum of a Narrow Line Seyfert 1 galaxy PG 1448+273 observed with the \xmm\ observatory. The signatures of the outflow are observed in both hard X-rays (iron K band) and soft X-rays (0.3-2.0 keV), and we constrain its physical properties systematically using an automated multi-parameter search of the X-ray spectrum with photo-ionised absorption grids.

PG 1448+273 is a Narrow Line Seyfert 1 AGN \citep{Boller+96, Grupe+04} located at a redshift of z=0.0645 \citep{Alam+15}. Its mass is $(9 \pm 2) \times 10^{6} M_{\odot}$ \citep[][from single-epoch optical spectroscopy]{Vestergaard+06}. PG 1448+273 has an estimated bolometric luminosity of $\sim 10^{45.5}$ erg/s \citep{Grupe+04}, resulting in an Eddington ratio above unity ($L/L_{\rm Edd}\sim3$). PG 1448+273 could therefore be a super-Eddington accretor \citep{Kawaguchi+03}. At such a high Eddington fraction, radiation pressure is an important component of the accretion flow and is expected to drive powerful outflows from the accretion disc \citep{Shakura+73}. PG 1448+273 is therefore a natural candidate for a spectroscopic search of UFO signatures.

Previous X-ray studies have remarked on the complex shape of the soft X-ray excess \citep{Innoue+07} of PG 1448+273, requiring more than the usual $\sim$0.1 keV blackbody component for an accurate description. The AGN is however not heavily absorbed by neutral absorption or a strong warm absorber. Its X-ray spectrum is also highly variable \citep{Ponti+12b}.

This paper consists of 5 sections. The observation and data reduction is described in Section \ref{data}. Details of our analysis and the results are shown in Section \ref{results}, followed by a discussion of the findings in Section \ref{discussion}. Section \ref{conclusions} contains the conclusions of this study.

\section{Observations and Data Reduction}
\label{data}

The \xmm\ \citep{Jansen+01} data were obtained from the XSA\footnote{nxsa.esac.esa.int/nxsa-web/} archive. The main dataset consists of a single observation taken on 24/01/2017 (ID: 0781430101, proposal PI: A. Zoghbi) with a total raw exposure of 126 ks. There is also a second, much shorter (22 ks) observation, which was taken on 08/02/2003 (PI: Kawaguchi).

We reduced the main dataset using standard pipelines with SAS v17. There were no high background periods of time during the observation and thus no time intervals needed to be filtered out.

The observation with the European Photon Imaging Camera (EPIC) pn and MOS instruments was split into two exposures, a scheduled (`S') and an unscheduled (`U') part. Initially we extracted the spectra of the two exposures separately, but found no significant differences between them. We therefore combined the scheduled and unscheduled parts of the observation into a single exposure.

\begin{figure}
	\includegraphics[width=\columnwidth]{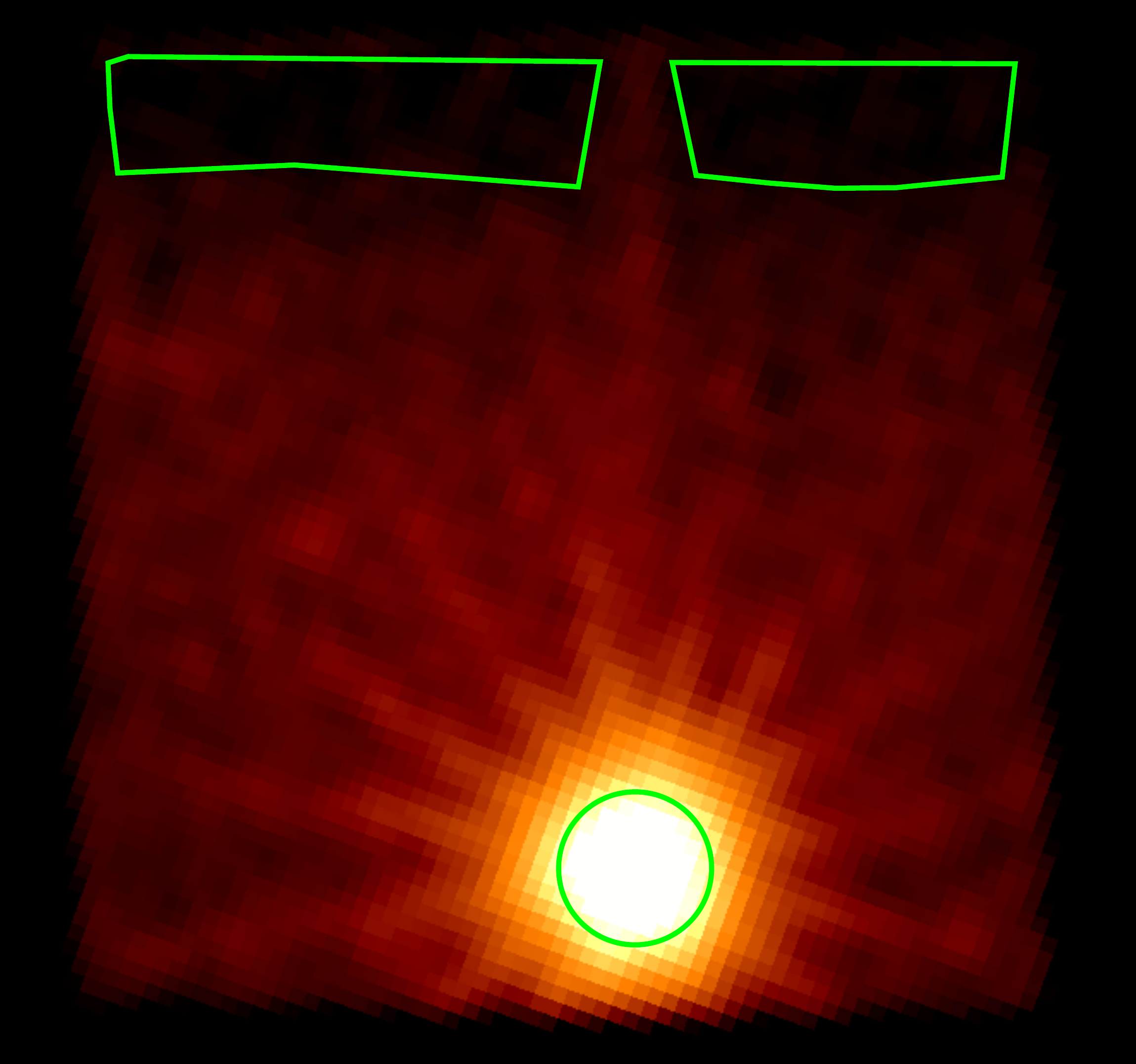}
    \caption{Image of the EPIC pn Small Window Mode exposure. The source region used in the analysis is the green circle (20 arcsec radius), the 2 polygons are the regions used to extract the background spectrum. The non-uniform background count distribution (in the vertical direction) is evident and is likely caused by the wings of the source point spread function. }
    \label{PN_image}
\end{figure}

The EPIC pn \citep{Struder+01} instrument was operated in Small Window mode during the observation to limit pile-up. An image of the pointing was made using standard routines. We noticed a strong non-uniformity of background counts across the image (Fig. \ref{PN_image}), likely due to source counts from the wing of the source point spread function (PSF). To avoid including source counts in the background spectrum, we chose background regions to be two small polygons as far from the source as possible. Furthermore, we decreased the importance of background subtraction by choosing a small (20 arcsec) source region. This boosted the signal-to-noise ratio in the critical $6-10$ keV energy band where the Doppler-shifted absorption features of highly ionised iron (\ion{Fe}{XXV/XXVI}) are expected. Only events of PATTERN <= 4 (single/double) were accepted. The average EPIC pn count rate was 2.2 ct/s, but the source is strongly variable, varying between 1 ct/s and 4 ct/s over the course of the observation.

EPIC/MOS 1 and 2 \citep{Turner+01} were operated in Large Window mode. The source region was chosen to be a circle with a radius of 20 arcsec, similar to pn data. The background covered the rest of the central chip area, at least 110 arcsec away from the source, avoiding other bright point sources. We did not notice any background non-uniformity as observed in EPIC pn data, confirming that the non-uniformity observed in the EPIC pn image is caused by the wings of the PSF. Events of PATTERN <= 12 (single/double) were accepted. The average EPIC MOS 1 count rate was 0.44 ct/s and the average MOS 2 count rate was 0.43 ct/s. We stacked data from MOS 1 and MOS 2 into a single spectrum with the \textsc{epicspeccombine} routine.

Data from both EPIC/pn and EPIC/MOS were grouped using the \textsc{specgroup} procedure to at least 25 counts per bin, and to oversample the detector resolution by a factor of at most 3. The clean exposure time of the pn dataset is 76 ks, and the exposure of the combined MOS 1 and MOS 2 data is 108 ks (pn exposure is smaller due to a lower Live time in Small Window mode).

Reflection Grating Spectrometer \citep[RGS,][]{denHerder+01} data were reduced using standard routines with default values and with standard observational background. Data were binned by a factor of 3 directly within the SPEX fitting package to oversample the real grating resolution by roughly a factor of 3. The clean exposure of RGS 1 and RGS 2 data is 115 ks.

After reduction, the data were converted into \textsc{spex fits} format using the \textsc{trafo} tool. \textsc{spex} fitting package \citep{Kaastra+96} was used for spectral fitting as it contains the physically-motivated ionised absorption grids \textsc{xabs} and \textsc{pion}. All spectral models were fitted using Cash statistics \citep{Cash+76} and the errors are stated at 1$\sigma$ level. We use 3 cross-calibration constants to account for calibration differences between the four spectral datasets (pn, MOS, RGS1 and RGS2). The differences are smaller than 10 per cent in all cases.

EPIC pn and MOS data were initially ignored below 0.3 keV and above 10 keV following standard guidelines. RGS data were used in the range between 7.5 \AA\ (1.7 keV) and 28 \AA\ (0.44 keV), limited by strong background on both ends of this interval. We found a 10-15 per cent discrepancy between EPIC and RGS data in the 0.3-1.0 keV range which could be caused by poor spectral resolution of the EPIC instruments in this range, or by calibration differences. At these energies EPIC lacks the resolution to resolve the fine structure of absorption lines but would drive the spectral fits due to much higher count rates than RGS (resulting in much smaller errorbars compared to RGS data). To avoid confusion between different unresolved spectral components, we ignored EPIC pn and MOS data below 1.7 keV altogether in the main part of our analysis.

We followed the same steps to reduce and prepare the second observation (0152660101) for analysis. The statistics were naturally much worse compared to the main dataset, albeit the AGN was brighter in 2003. The average EPIC pn count rate was 4 ct/s, and the average MOS 1/2 count rate was 0.9 ct/s. Since EPIC pn was operated in Large Window mode (with a pile-up limit of 3 ct/s), it is possible that the pn data are affected by pile-up. MOS was operated in Large Window mode (pile-up limit of 1.5 ct/s) and its data should not be affected. Nevertheless, we do not observe any significant differences between pn and MOS spectra. We ignored EPIC data below 1.7 keV, only using RGS data in the soft X-ray band, following the same procedure as with the main dataset.

\begin{figure}
	\includegraphics[width=\columnwidth]{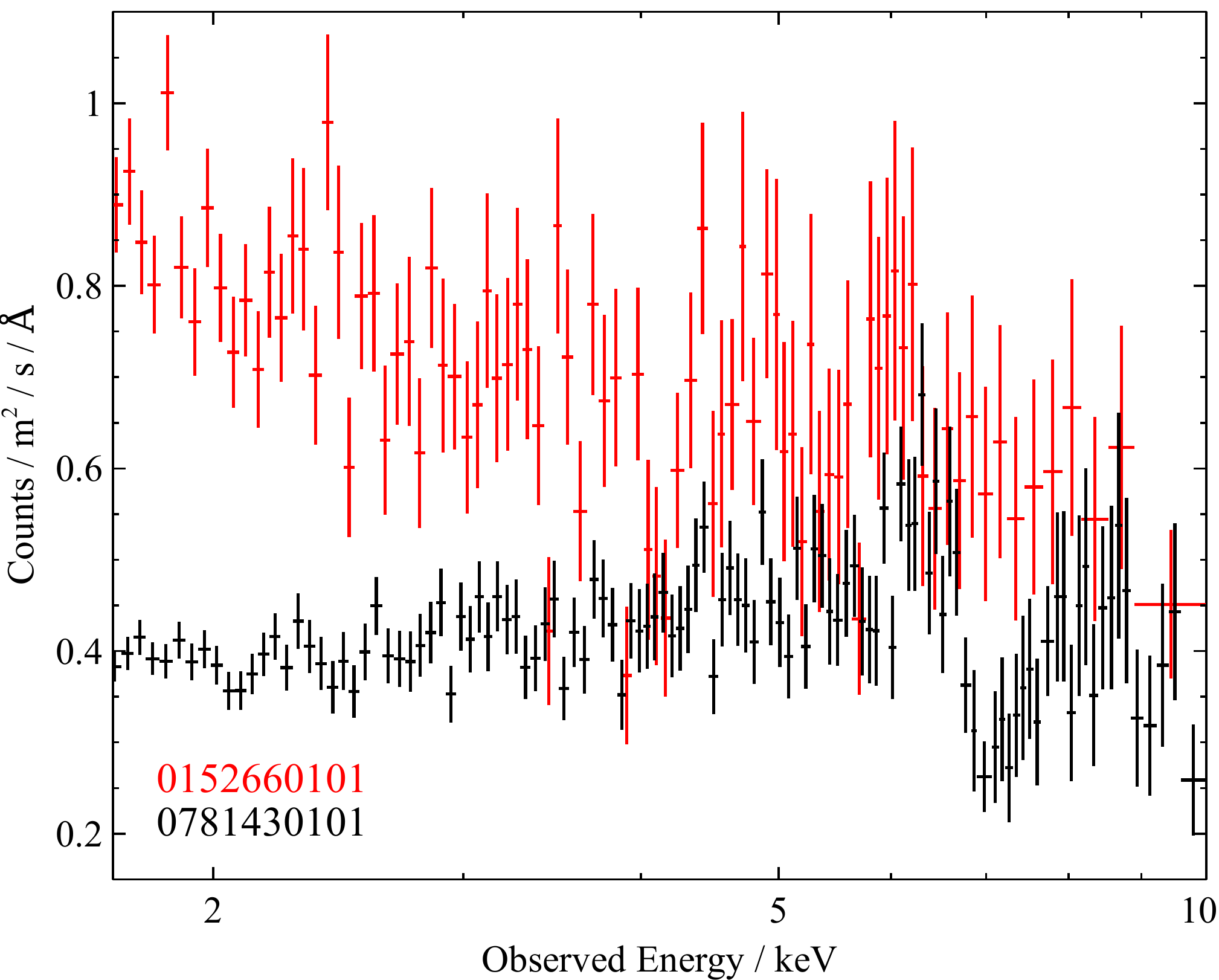}
    \caption{$1.7-10$ keV EPIC pn spectra of PG 1448+273 during the two observations. Data from the main observation (0781430101) are in black colour, data from the shorter archival observation (0152660101) are shown in red.}
    \label{PN_015_078}
\end{figure}

Fig. \ref{PN_015_078} contains the $1.7-10$ keV EPIC pn spectra from both observations and shows the longterm variation of the source and the dataset quality. Especially striking is the difference in the iron K region ($5-8$ keV). We first focus on the analysis of the higher quality observation (0781430101) which also shows more intriguing spectral features. The second observation is analysed separately in Sect. \ref{oldobs}. It would not be appropriate to stack the two datasets given the long period of time and strong variability between the two observations.

\section{Results from the main \xmm\ observation}
\label{results}

\subsection{Broadband X-ray continuum}
\label{continuummodelling}

First we fit the broadband 0.3-10 keV spectrum with a phenomenological X-ray AGN spectral model. We fit the RGS, EPIC pn and MOS data simultaneously with the same spectral model except for the cross-calibration constant. The model consists of a powerlaw (\textsc{pow} in SPEX) model describing the Inverse Compton coronal emission, a blackbody (\textsc{bb}) model describing the soft excess below 2 keV, and a relativistically blurred iron K emission line. Blurred reflection is described with a \textsc{laor} model applied to a Gaussian line (\textsc{ga}) with the energy fixed to 6.4 keV to avoid too much model freedom with the available data quality. All of these models are subject to any interstellar absorption in the host galaxy of the AGN described by a \textsc{hot} model within SPEX. The model is then redshifted by z=0.0645 of the host galaxy (found on the NED database) using the \textsc{reds} model. Finally, Galactic interstellar absorption is applied with a second \textsc{hot} spectral model with a fixed value of $N_{\textrm{H}}=$2.5$\times10^{20}$ \pcm\ \citep{Kalberla+05}. The final spectral model has the following form: \textsc{hot(reds(hot(pow+bb+laor$\times$ga)))}.

We find that PG 1448+273 shows properties very similar to other Narrow Line Seyfert 1 AGN (for a summary of the continuum parameters, see Table \ref{Contparams}). The powerlaw slope is measured to be $2.07 \pm 0.03$, and the soft excess can be described with a blackbody of temperature $0.097 \pm 0.003$ keV. The black hole appears to have a high spin value with a minimum disc radius of just $1.52_{-0.29}^{+0.19}$ R$_{\textrm{G}}$, but we draw no firm conclusions since the relativistic iron K reflection is described only with a simple phenomenological model. The emissivity slope is q$=4.6 \pm 0.5$ and the measured disc inclination $44_{-3}^{+2}$ degrees. We also find a significant non-zero interstellar absorption in the host galaxy of PG 1448+273 with a column density of $(9.8 \pm 2.1) \times 10^{20}$ \pcm. The fit statistics are C-stat=1594.43 for 1278 degrees of freedom (D.o.F.).

\begin{figure}
	\includegraphics[width=\columnwidth]{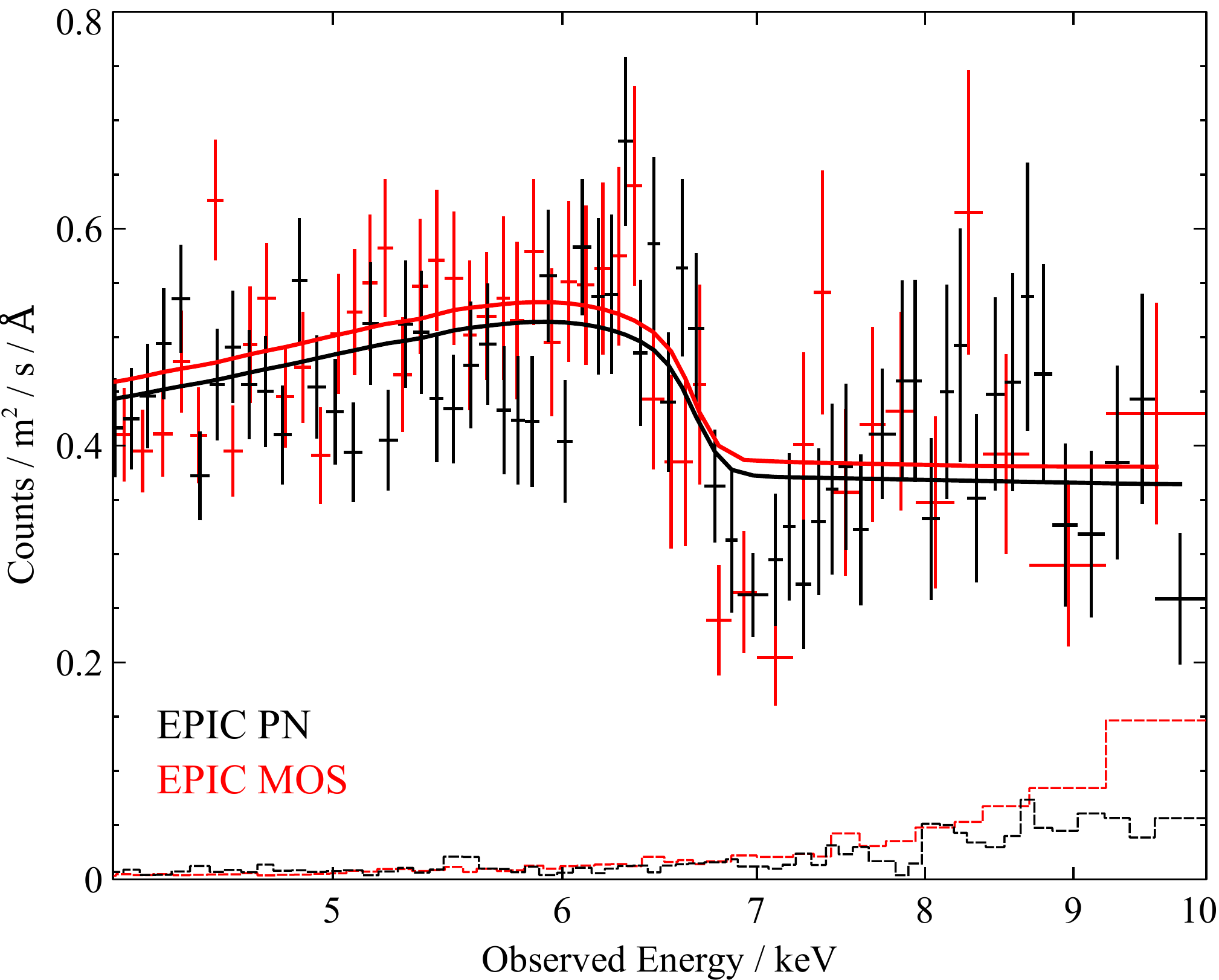}
    \caption{EPIC pn (black) and MOS (red, MOS 1 and MOS 2 stacked) data ($4.5-10$ keV) of PG 1448+273 fitted with the phenomenological AGN continuum model from Section \ref{continuummodelling}. Apparent is the strong absorption residual observed at $\sim$7 keV. Black (pn) and red (MOS) solid lines show the best-fitting continuum models, and the dashed lines show the subtracted background spectra.}
    \label{PN_spectrum}
\end{figure}

\begin{figure*}
	\includegraphics[width=\textwidth]{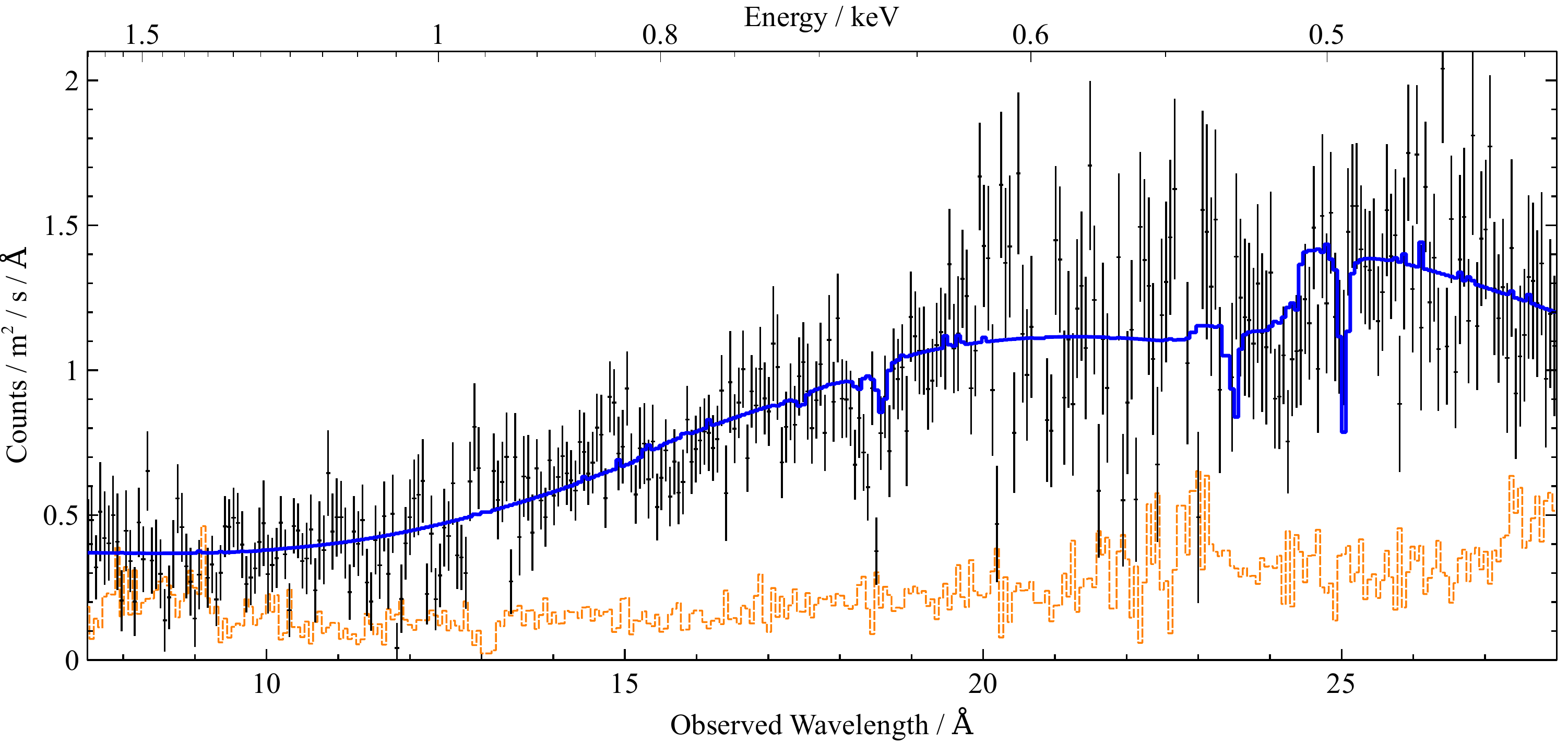}
    \caption{Background-subtracted RGS data (7.5-28 \AA) of PG 1448+273, stacked and overbinned for plotting purposes only, fitted with the phenomenological AGN continuum model from Section \ref{continuummodelling} (blue colour). The model particularly fails to reproduce the strong residuals around 18-19 \AA, however further residuals are seen around 12, 15.5 and 24 \AA. The edge at 24.5 \AA\ and the strong absorption line at 25 \AA\ suggest the presence of neutral absorption intrinsic to the AGN. The subtracted background is shown in orange.}
    \label{RGS_spectrum}
\end{figure*}

Upon visual inspection of the spectrum, strong residuals in EPIC pn and MOS data within the iron K band are obviously evident (Fig. \ref{PN_spectrum}), as are residuals in the RGS band (Fig. \ref{RGS_spectrum}). The iron K band residuals suggest the presence of a highly ionised, high-velocity outflow, the flux drop above 7 keV is very similar to that of another NLS1 1H 0707$-$495 with a known high-velocity outflow \citep{Boller+02,Kosec+18c}. The RGS data point to a more complex situation. There is a prominent absorption residual at 18-19 \AA\ (observed wavelength), which could correspond to \ion{O}{VIII} absorption, blueshifted by $\sim$0.1c. Further structures are also seen around 12 \AA\ and 24 \AA\ in observed wavelength.

\subsection{Ionised wind search}
\label{wind_scan}

\subsubsection{Wind search set up and results}

To determine the blueshift and physical parameters of any ionised outflow component as well as its significance in the present data, we systematically scan the X-ray spectrum of PG 1448+273 with an automated ionised absorber search. In principle, we use a routine to scan a large parameter space of a photoionisation absorber grid, to locate the best-fitting ionised outflow properties. Similar methods were recently successfully used to detect and study ultrafast outflows in Ultraluminous X-ray Sources \citep{Kosec+18a, Kosec+18b, Pinto+20}.

The photoionisation model we use here to describe the spectral features (absorption lines) of an ionised outflow is called \textsc{xabs} in the SPEX fitting package. The model calculates the ionising balance and absorption line strengths based on the Spectral Energy Distribution (SED) shape of the AGN NGC 5548 \citep{Steenbrugge+03, Steenbrugge+05}. The SED of NGC 5548 should be a reasonable approximation for the PG 1448+273 SED, although the former is likely more massive at $(9 \pm 3) \times 10^7 M_{\odot}$, and has a smaller Eddington ratio \citep{Lu+16}. The coronal powerlaw slope of NGC 5548 is $\Gamma\approx1.8$ versus $\Gamma=2.07 \pm 0.03$ in PG 1448+273, so PG 1448+273 has a softer spectrum but not as soft as the more extreme NLS1 AGN like IRAS 13224$-$3809 where $\Gamma$ varies between 2.2 and 2.9 \citep{Pinto+18}. At each step, we also check our results for consistency with the \textsc{pion} model in the \textsc{spex} fitting package. \textsc{pion} is a similar model of photo-ionised plasma as \textsc{xabs} but takes into account the actual continuum SED of the currently used spectral model, albeit at a much higher computational cost, which is the limiting factor in our analysis.

The main parameters of the \textsc{xabs} model are the column density $N_{\textrm{H}}$, ionisation parameter $\log \xi$, systematic velocity $z$ and turbulent velocity (or simply velocity broadening) $v_{\textrm{turb}}$ of the ionised absorber. An outflow in the X-ray spectrum of the AGN can in principle have any blueshift, ionisation parameter, column density and turbulent velocity. This is a large parameter range which has to be searched in an automated way to avoid missing the parameter space with the best-fitting solution. As an added bonus, performing a systematic search in this way will allow us to constrain the statistical significance of any outflow detection in the AGN spectrum.

We therefore create a grid of photoionisation models which spans ionisation parameters between $\log (\xi/ \textrm{erg cm s}^{-1})=-1.0$ and $\log (\xi/ \textrm{erg cm s}^{-1})=4.8$ with a grid spacing of $\Delta\log\xi=0.2$ \citep[in line with the recommendation of][]{Reynolds+12}, with three fixed turbulent velocities of $v_{\textrm{turb}}=100$ km/s, $v_{\textrm{turb}}=1000$ km/s and $v_{\textrm{turb}}=10000$ km/s. The turbulent velocity space grid is chosen to be coarse to decrease the computational cost of the automated search.

Afterwards a range of realistic systematic absorber velocities (blueshifts) is chosen. We choose to search for an absorber with a systematic velocity between +20000 km/s (to account for possible infalling gas) and -100000 km/s (blueshift of roughly 0.4c, after relativistic correction). The step size is 150 km/s for turbulent broadening of 100 km/s and 300 km/s for turbulent broadening of 1000 km/s. This is chosen so that in each case the step size is not much larger than the broadening and at the same time is not much larger than the spectral resolution of our detectors, RGS being the best of our instruments with a spectral resolution of roughly $\sim$300 km/s in velocity units. In the case of turbulent velocity equal to 10000 km/s, the features are so broad that such detailed sampling is not necessary, instead we sample by 1000 km/s.

We generate an absorption grid for every possible systematic velocity, ionisation parameter and broadening within the selected range, add this model to the baseline continuum from Section \ref{continuummodelling}, and fit for the column density of the absorber (with the remaining absorber parameters fixed), leaving the continuum parameters free to vary. If an absorber of such parameters is disfavoured by the data, the resulting column density will be 0, i.e. the best-fitting solution is the null (baseline continuum) model. Otherwise we recover the best-fitting absorber column density as well as the \delcstat\ fit improvement obtained by adding the photoionisation grid to the (null) baseline continuum model. Any improvement in the \delcstat\ fit statistics compared to the null model is recorded for every point in the multi-dimensional search grid, and determines how strongly the spectral model including the wind absorption is preferred to the baseline continuum model.

\begin{figure*}
	\includegraphics[width=\textwidth]{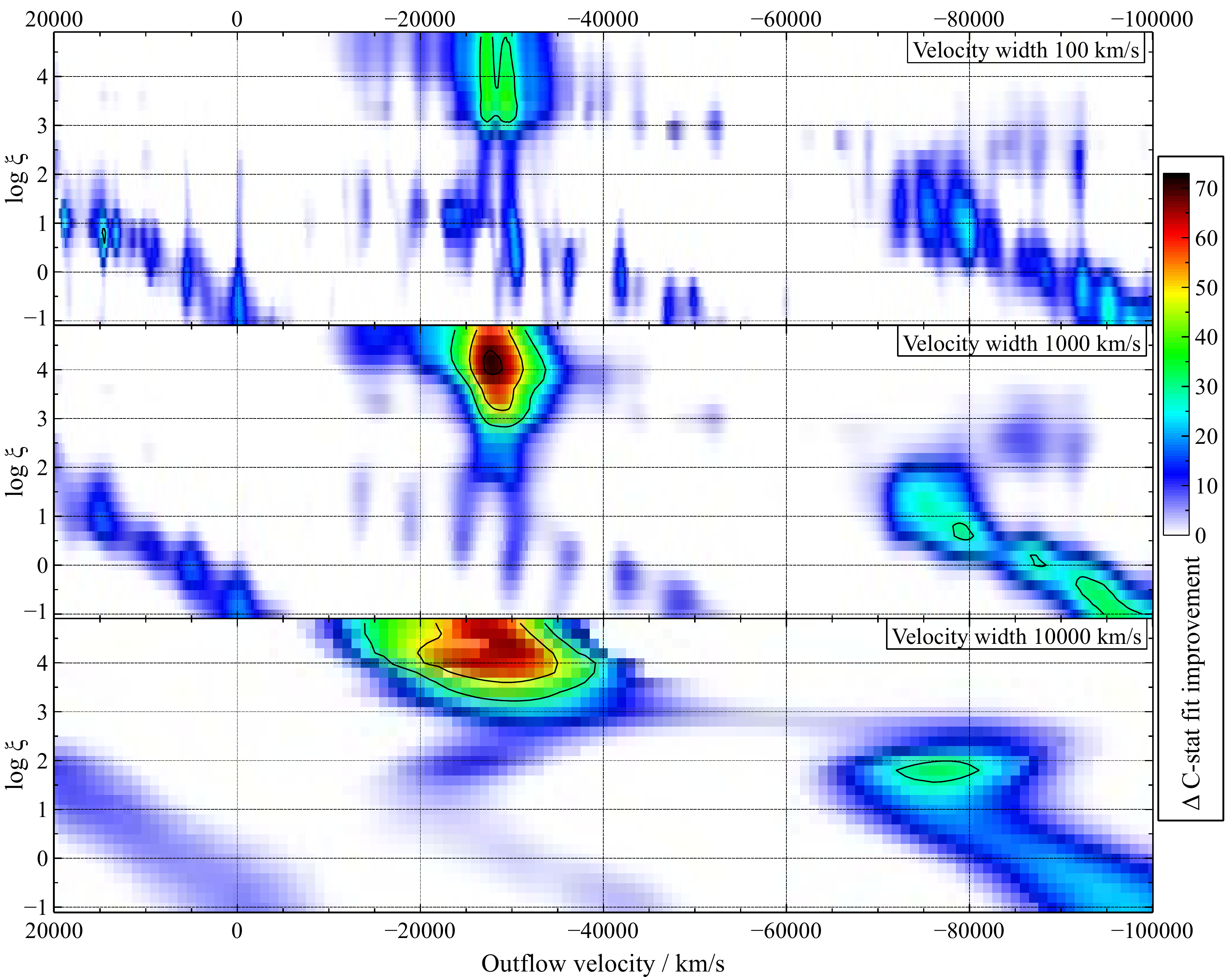}
    \caption{Systematic scan for ionised outflow features in the spectrum of PG 1448+273. The spectrum is scanned with ionised absorption model grids with systematic velocities between +20000 km/s and -100000 km/s (on the X axis), ionisation parameters $\log \xi$ between -1.0 and +4.8 (on the Y axis), with turbulent velocities of 100 km/s (top subplot), 1000 km/s (middle subplot) and 10000 km/s (bottom subplot). The colour shows the \delcstat\ fit improvement upon adding the absorption grid to the baseline continuum model (according to the colour bar on the right), the black contours show \delcstat\ fit improvement of 30, 50 and 70, respectively.}
    \label{2D_Scan_cont}
\end{figure*}

We search the large parameter space for signatures of an ionised outflow as described above, using the full dataset (RGS + pn + MOS) on PG 1448+273. The results of the search are shown in Fig. \ref{2D_Scan_cont}. The wind search achieves a strong detection of ionised plasma with a systematic outflow velocity of $\sim30000$ km/s, with a fit improvement of \delcstat$>$70. The plasma is highly ionised with $\log \xi$ of $4.0-4.2$ and the fit improvement is much higher in the scans with a 1000 or 10000 km/s turbulent velocity compared to the one with 100 km/s velocity width. All the other \delcstat\ peaks in the systematic scans are much smaller, reaching at most \delcstat$\sim$30-35, hence the `primary' (strongest) solution is highly preferred.

The fact that our detection is above \delcstat=70 already strongly suggests the significance of this outflow component, comparing with previous similar analyses \citep[e.g.][]{Kosec+18b}. However, due to the look-elsewhere effect, since we are searching through a large parameter space, it is not trivial to assign directly a `wind detection significance' to a specific value of \delcstat\ \citep{Protassov+02}. Monte Carlo simulations must be employed to rigorously `map' the C-stat fit improvement to a specific false alarm probability for each wind detection. The results of Monte Carlo simulations are shown in Appendix \ref{MCSims}. They indicate that the significance of the primary peak detection is well above $3-4\sigma$ and thus our UFO detection is robust.

\subsubsection{Secondary detections in the wind search}

\begin{figure*}
	\includegraphics[width=\textwidth]{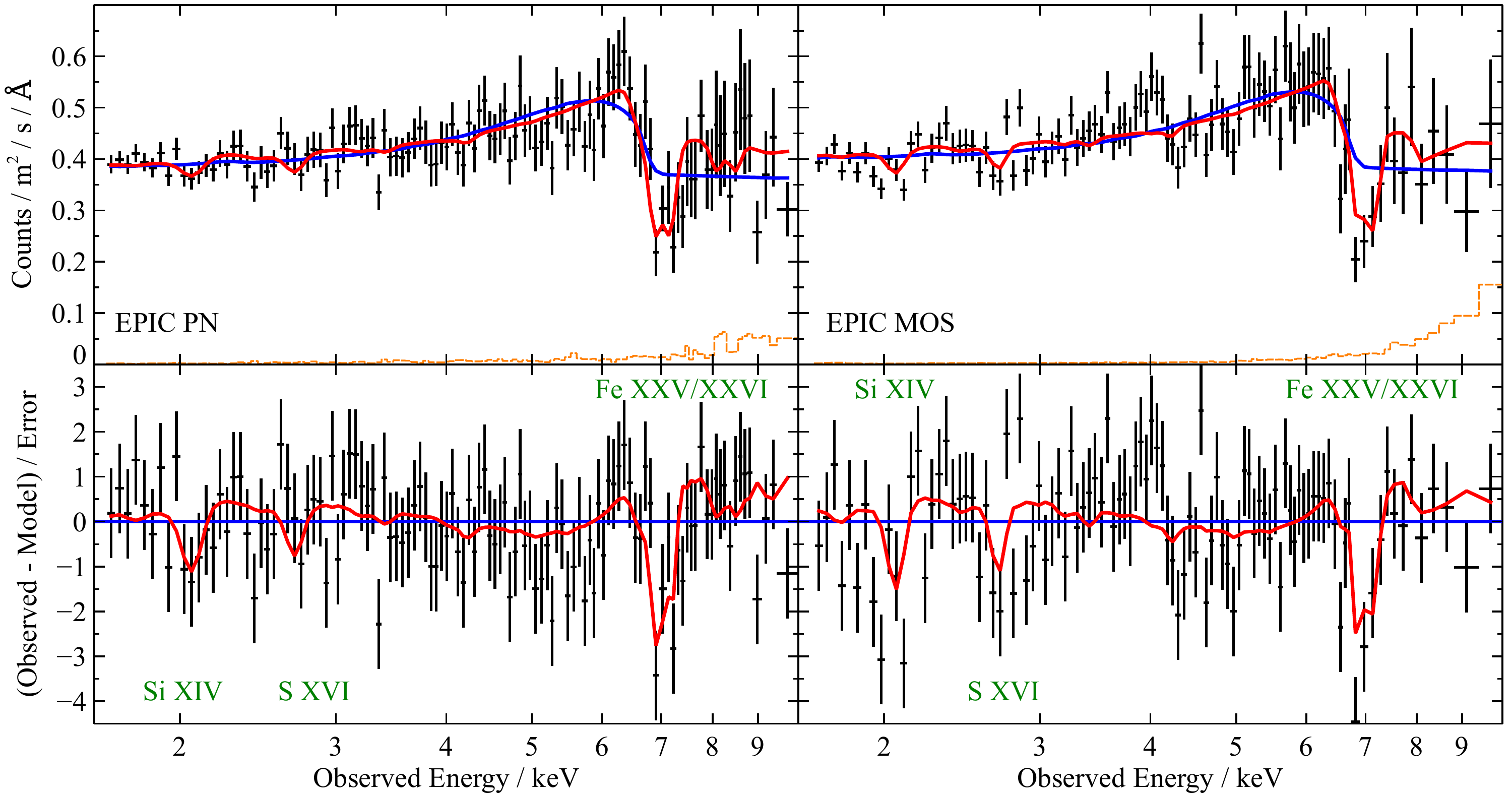}
    \caption{EPIC pn (left subplots) and MOS (right subplots) data ($1.7-10$ keV) of PG 1448+273 fitted with the phenomenological AGN continuum model from Sect. \ref{continuummodelling}, plus two wind components: one ultrafast, highly ionised component outflowing at $\sim$27000 km/s, and a warm absorber component in the rest-frame of the AGN. The top subplots show the spectra with the best-fitting baseline continuum (blue colour) and the continuum + wind models (red colour), the bottom subplots contain the residuals to the continuum model as well as the best-fitting wind solution (red). The orange dashed lines show the subtracted background spectra. Green labels name the strongest absorption lines of the UFO absorption.} 
    \label{PN_xabs_warm}
\end{figure*}

\begin{figure*}
	\includegraphics[width=\textwidth]{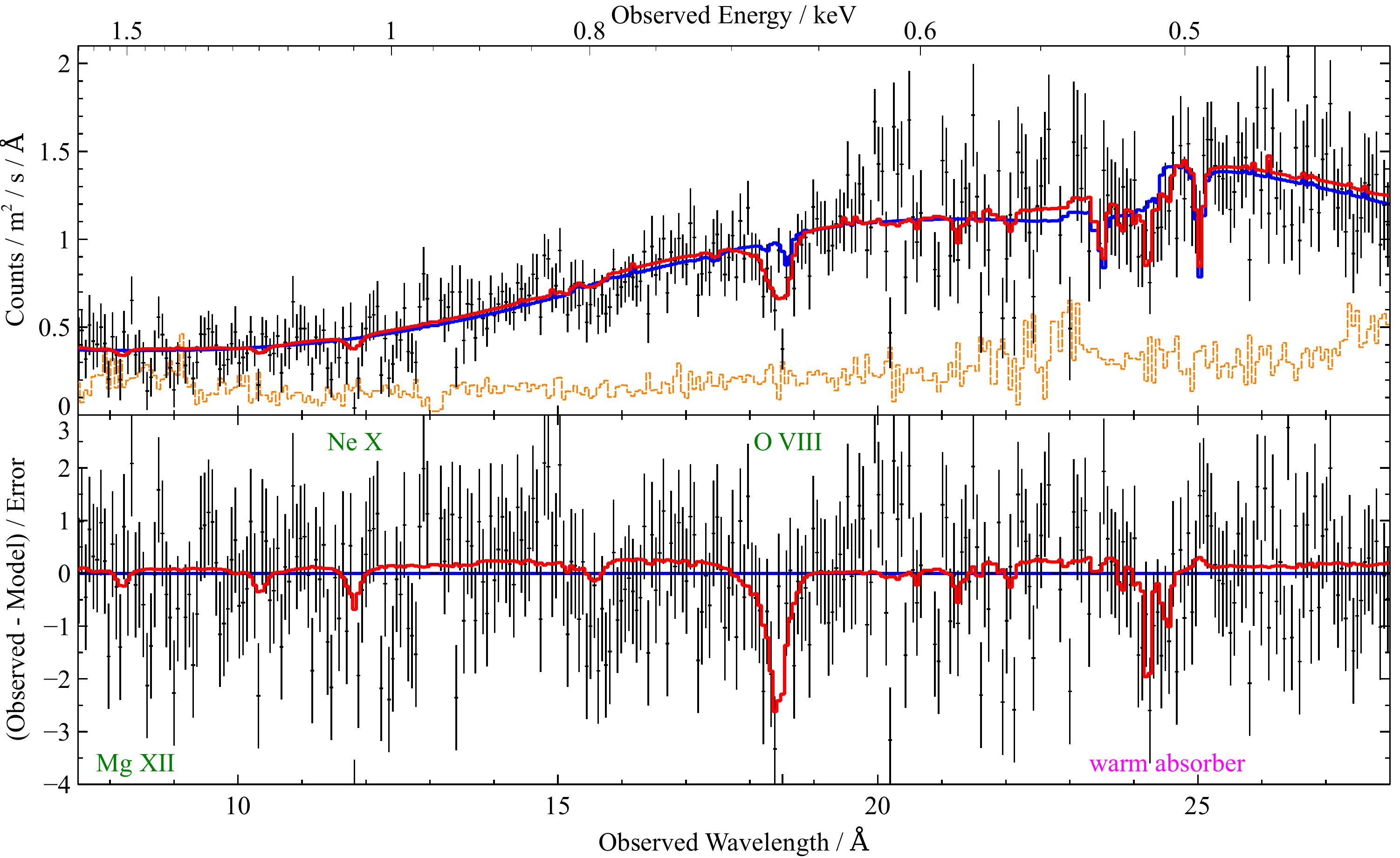}
    \caption{RGS data between 7.5 \AA\ (1.7 keV) and 28 \AA\ (0.4 keV) of PG 1448+273, stacked and overbinned for plotting purposes only, fitted with the phenomenological AGN continuum model from Section \ref{continuummodelling}, plus two ionised components: one ultrafast, highly ionised component moving at $\sim$27000 km/s, and a warm absorber component in the rest-frame of the AGN. The top subplot contains the RGS spectrum fitted with the baseline continuum (blue colour) and continuum + wind (red colour) models, the bottom subplot shows the residuals to the continuum model and the best-fitting wind solution (red). The subtracted background in orange colour. Green labels name the most notable absorption lines of the UFO absorption and the magenta label shows the spectral features of the warm absorber.}
    \label{RGS_xabs_warm}
\end{figure*}

We also find a large number of secondary peaks with \delcstat\ between 25-35 which by themselves could be considered significant detections, if they were detected alone. Their presence suggests that there could be further wind structure hiding in the spectrum of PG 1448+273. However, at this point it is not certain that the secondary peaks are independent of the primary peak. These secondary solutions could in fact be fitting some of the same spectral residuals to the baseline continuum as the primary peak, but with a different set of atomic lines, resulting in a shallower \delcstat\ improvement.

Finally, we also notice a \delcstat\ peak around the rest-frame of the AGN in the 100 km/s velocity width search. The peak is not particularly strong with \delcstat$\sim$20 at the ionisation parameter of $\log \xi = -0.6$, but could mean the presence of a weak warm absorber. Absorbers with similar ionisation parameters are very commonly observed in other AGN \citep[e.g.][]{Blustin+05}, it is thus not unlikely than one could also appear in the spectrum of PG 1448+273. If this is the case, the same look-elsewhere effect as for the UFO detection does not apply. Warm absorbers in particular only appear over a very small systematic velocity parameter space (velocities from hundreds of km/s to a few thousands of km/s at maximum). The same \delcstat\ significance rules for UFO detections hence do not apply as the available parameter space is much smaller, and a significantly weaker warm absorber detection might still represent real absorption features. For this reason we fit both the features of the UFO primary peak as well those of the warm absorber in the next section.

\subsection{Ultrafast outflow and warm absorption}
\label{Sec_xabs_warm}

To learn more about the properties of the ionised absorbers in PG 1448+273, we fit its spectrum directly with a model which includes the original baseline continuum (with all the parameters freed), plus two ionised absorbers described with the \textsc{xabs} model. The first \textsc{xabs} model is supposed to fit the primary peak in the systematic absorber search at around 30000 km/s, the second one should fit the possible features of a warm absorber in the rest-frame of the AGN.

The primary ultrafast component is found with an outflow velocity of $26900 \pm 600$ km/s (including the relativistic correction). Its column density is $0.28_{-0.07}^{+0.12} \times 10^{24}$ \pcm, with an ionisation parameter of $\log (\xi / \textrm{erg cm s}^{-1})=4.03_{-0.08}^{+0.10}$ and a velocity width of $2100_{-500}^{+600}$ km/s. The final fit improvement of adding this component to the baseline model is very strong with \delcstat=83.27, confirming the significance of the ultrafast outflow features in the spectrum of PG 1448+273. The strongest spectral features of the UFO absorption are the \ion{Fe}{XXV} and \ion{Fe}{XXVI} lines at $\sim$7 keV (in the EPIC band) and the \ion{O}{VIII} line at 18.5 \AA\ ($\sim$0.7 keV, in the RGS band).

The warm absorber has a column density of $(3.8 \pm 1.7) \times 10^{20}$ \pcm\ and an ionisation parameter of $\log (\xi / \textrm{erg cm s}^{-1})=-0.4_{-0.4}^{+0.3}$. Its systematic velocity is $-100 \pm 200$ km/s in the rest-frame of the AGN so it is consistent with being at rest, with a velocity width of $50_{-30}^{+40}$ km/s. It is similar to warm absorbers found in other AGN \citep{Kaspi+02, Steenbrugge+05, Detmers+11}, apart from the column density which is much lower in PG 1448+273, resulting in rather weak spectral features, and a lower \delcstat\ fit improvement of 15.19. The \delcstat\ is smaller than the original value found in the automated wind search ($\sim$20), suggesting that the ultra-fast and warm absorber models are partly fitting the same residuals to the original baseline continuum. The spectral features of the warm absorber are located at around 24 \AA\ in the RGS band.

Since the strongest absorption lines of the warm absorption lie in the 24 \AA\ range, they overlap with the \ion{N}{VII} absorption of the 0.1c ultra-fast phase. It is therefore possible that these features instead could originate in \ion{N}{VII} absorption of the ultra-fast phase. We attempt to describe the features by freeing the abundance of N in our final spectral model (with the warm absorber column density fixed to 0), and reach fit statistic improvements at a \delcstat\ $\approx10$ level (compared to a continuum + UFO with solar abundances spectral model), for a nitrogen over-abundance of $7^{+3}_{-2}$. We consider this to be a very high abundance value and prefer the warm absorption origin for the 24 \AA\ features as a far more likely explanation.

The baseline continuum parameters slightly shift (for a summary of the continuum parameters, see Table \ref{Contparams}), with the best-fitting powerlaw slope being $1.97 \pm 0.03$ and the soft excess blackbody temperature of $0.102 \pm 0.003$ keV, however they still agree with the original continuum fit values within $\sim3\sigma$ errors. The inner disc radius (from the iron K line) is $1.7 \pm 0.5$ R$_{\rm G}$, the emissivity slope is $3.5_{-0.5}^{+2.1}$ and the disc inclination is found to be $38_{-3}^{+16}$ degrees. Most notably, the best-fitting neutral absorber (in the rest-frame of PG 1448+273) is now weaker with a column density of $6.5_{-2.2}^{+2.0} \times 10^{20}$ \pcm. This could be explained if the warm absorber (with a low ionisation parameter) is fitting similar spectral residuals as the neutral absorption. The final fit statistics are C-stat=1495.97 for 1270 D.o.F.

The best-fitting wind spectral model is shown in Fig. \ref{PN_xabs_warm} (EPIC pn and MOS data only) and in Fig. \ref{RGS_xabs_warm} (RGS data only).

\subsection{Residual structure in the X-ray spectrum - a multiphase outflow?}

\begin{figure*}
	\includegraphics[width=\textwidth]{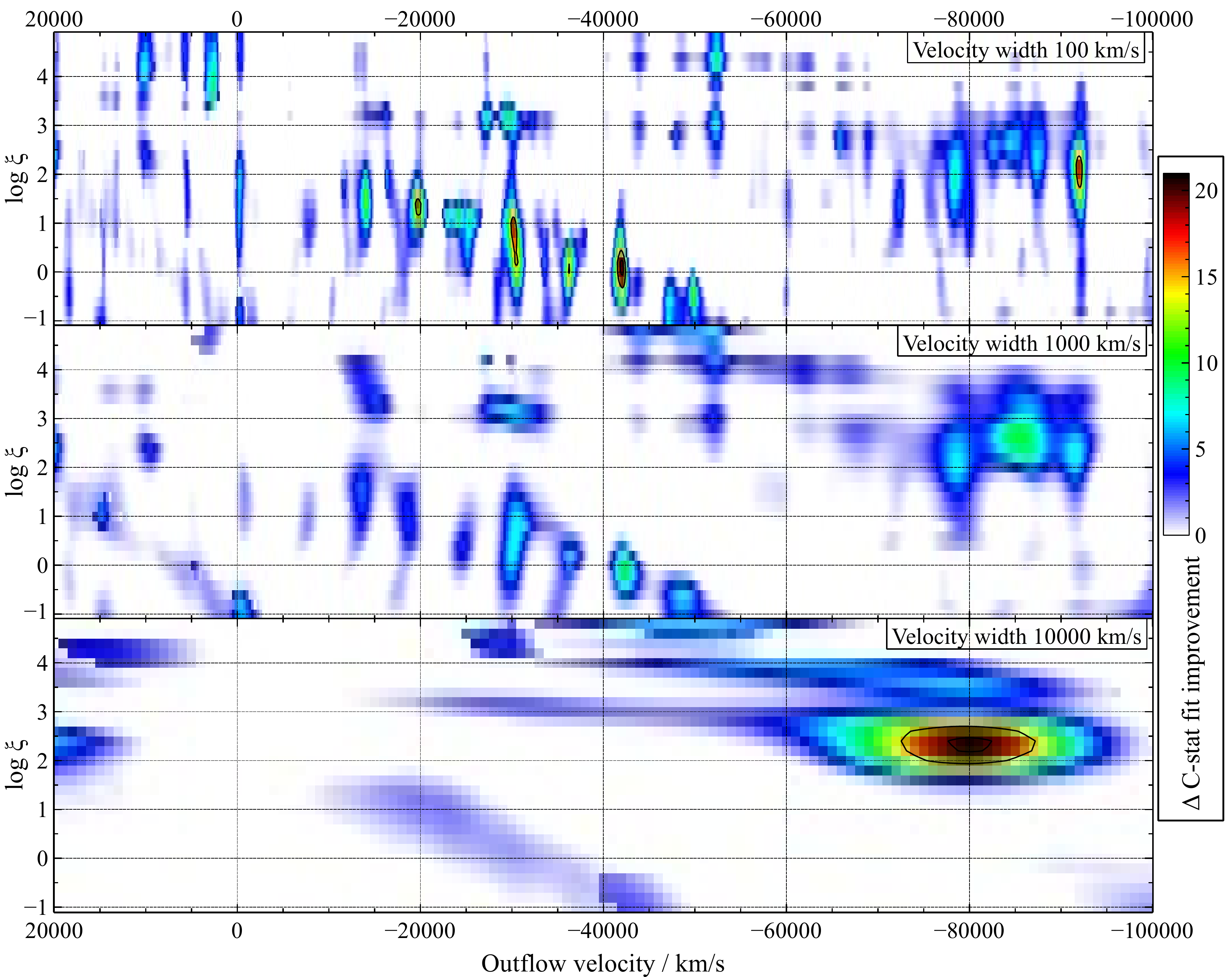}
    \caption{Systematic scan for ionised outflow features in the spectrum of PG 1448+273, in addition to the primary ultrafast component outflowing at 27000 km/s and a warm absorber component in the AGN rest-frame. The spectrum is scanned with ionised absorption grids with systematic velocities between +20000 km/s and -100000 km/s (the X axis), ionisation parameters $\log \xi$ between -1.0 and +4.8 (the Y axis), with turbulent velocity 100 km/s (top subplot), 1000 km/s (middle subplot) and 10000 km/s (bottom subplot). The colour shows the fit improvement \delcstat\ upon adding the absorption grid to the continuum model (according to the colour bar on the right), the black contours show \delcstat\ fit improvement of 15 and 20, respectively.}
    \label{2D_Scan_xabs_warm}
\end{figure*}

The large C-stat value of the best-fitting spectral model given the number of degrees of freedom, and the secondary peaks in the systematic wind search suggest the presence of further ionised outflow components in the spectrum of PG 1448+273, assuming our baseline continuum model is accurate. However, the secondary peaks might not be independent of the primary wind detection, i.e. their spectral signatures could be shared. For this reason, it does not make sense to directly fit the secondary peaks based on the existing automated wind search (Fig. \ref{2D_Scan_cont}).

We therefore perform a new set of wind searches which includes the two outflow components described above (in Sec. \ref{Sec_xabs_warm}). We choose the best-fitting baseline continuum with 2 additional \textsc{xabs} components (ultra-fast outflow and warm absorber) to be the new baseline model. We then run the same wind search as described in Sec. \ref{wind_scan}, with all the baseline continuum parameters allowed to vary freely. The results are shown in Fig. \ref{2D_Scan_xabs_warm}.

There are no strong wind detections in the search with a turbulent velocity of 1000 km/s. On the other hand, there are roughly 5 peaks with fit improvements between \delcstat\ of 15 and 22 in the search with 100 km/s turbulent velocity, suggesting tentative evidence of further ionised outflow components. Furthermore, there is a similarly strong peak (\delcstat$\sim$20) in the 10000 km/s width search. 

If only a single peak with a similar fit improvement (\delcstat$\sim$20) was detected, it would be trivial to assign a false alarm probability to its detection based on our Monte Carlo simulation results from Appendix \ref{MCSims}. However, the fact that there are 6 peaks with similar \delcstat\ values means that we likely cannot statistically differentiate which of these solutions are real and which are false detections as they will all result in similar statistical significances/false alarm probabilities. The situation is further complicated because these peaks are probably not independent, i.e. multiple potential peaks are fitting the same residuals to the baseline continuum, but with different outflow models.

We individually fit each of the peaks with \delcstat\ $> 15$ but find no further fit improvement or any particularly strong outliers. The tentative solutions mostly fit residuals in the RGS band, in particular those at 12-13 \AA, at 15-16 \AA\ and in the 20-22 \AA\ range (Fig. \ref{RGS_xabs_warm}). At the current data quality these residuals can be fitted equally well with multiple wind models. Since multiple wind solutions fit the RGS residuals similarly well, we cannot identify a unique solution with the present data. We find that none of the tentative peaks fit the remaining residuals seen in the iron K band, which could instead be interpreted as our imperfect fitting of the relativistic reflection with just the phenomenological Laor model (omitting the reflection continuum).

The large abundance of peaks with \delcstat\ of around 15-20 however still suggests a very likely presence of a multi-phase outflow in PG 1448+273 because such strong false detections should occur rarely by chance. Based on our MC simulations, even accounting for their limited parameter space, we can very roughly assign a 2$\sigma$ statistical significance (including the look-elsewhere effect) to a \delcstat=16 outflow detection. Such strong fake detections should therefore only occur in 5 per cent of observations, suggesting that the secondary peaks are real. At this moment, unfortunately, identification of these potential components is just beyond the possibilities of the current data quality.

\section{Archival short \xmm\ observation}
\label{oldobs}

We apply the same continuum model as used in Sect. \ref{continuummodelling} to the second, archival dataset. Due to lower data quality, we fix the blurred iron K line parameters to the best-fitting parameters from the final fit in Sect. \ref{Sec_xabs_warm}, which included the broadband continuum as well as the ultrafast and warm absorption. We find that the coronal powerlaw slope is much higher with $\Gamma = 2.38 \pm 0.04$, and the average 2-10 keV luminosity is $(2.16 \pm 0.07) \times 10^{43}$ erg/s.

\begin{figure*}
	\includegraphics[width=0.995\textwidth]{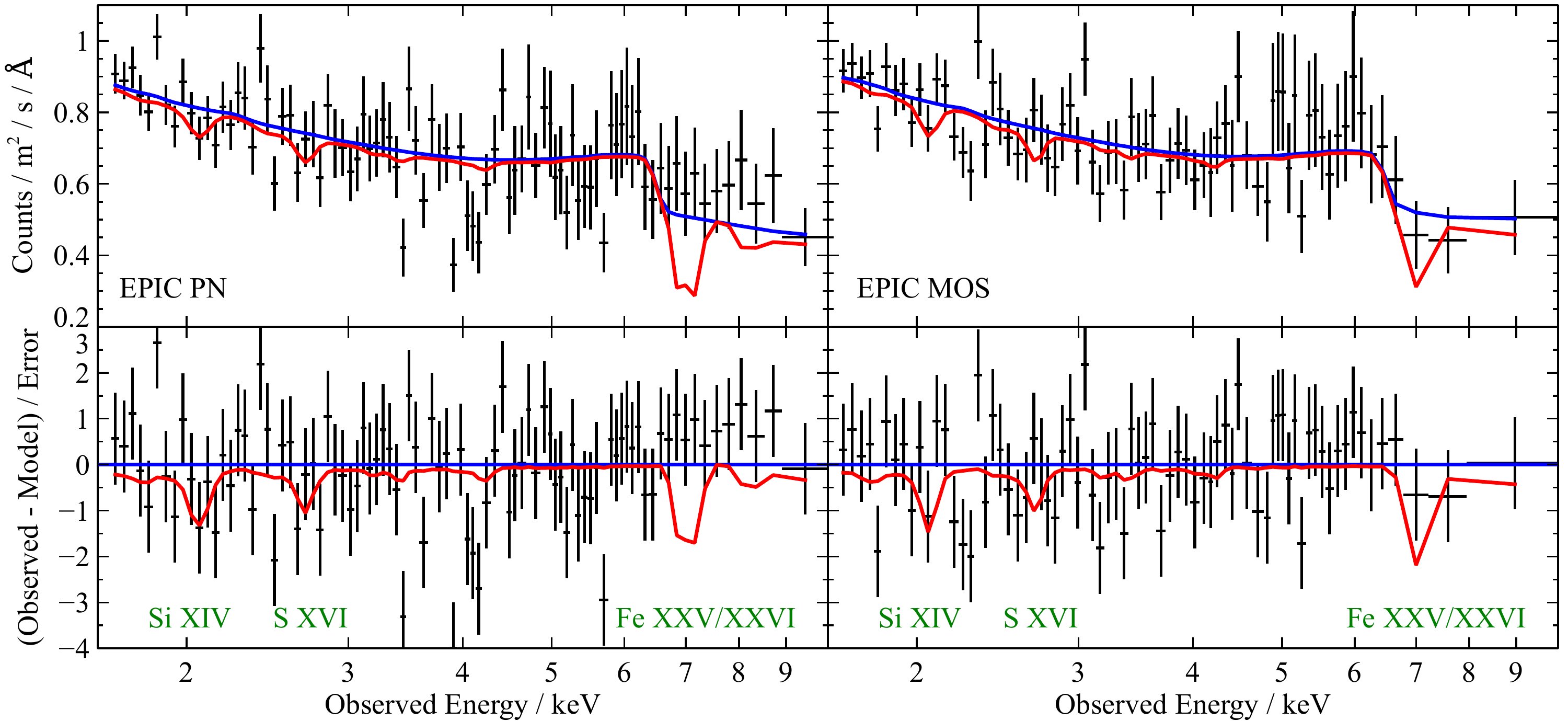}
    \caption{EPIC pn (left subplots) and MOS (right subplots) data ($1.7-10$ keV) of the second, short observation of PG 1448+273 fitted with the baseline continuum model from Appendix \ref{oldobs} (blue colour). The model in red colour is the best-fitting baseline continuum plus one UFO component with the parameters of the outflow detected in the long \xmm\ observation (Sect. \ref{Sec_xabs_warm}). The top subplots show the fitted spectra, the bottom subplots contain the residuals. Green labels name the strongest absorption lines of the UFO absorption.} 
    \label{PN_old}
\end{figure*}

\begin{figure*}
	\includegraphics[width=\textwidth]{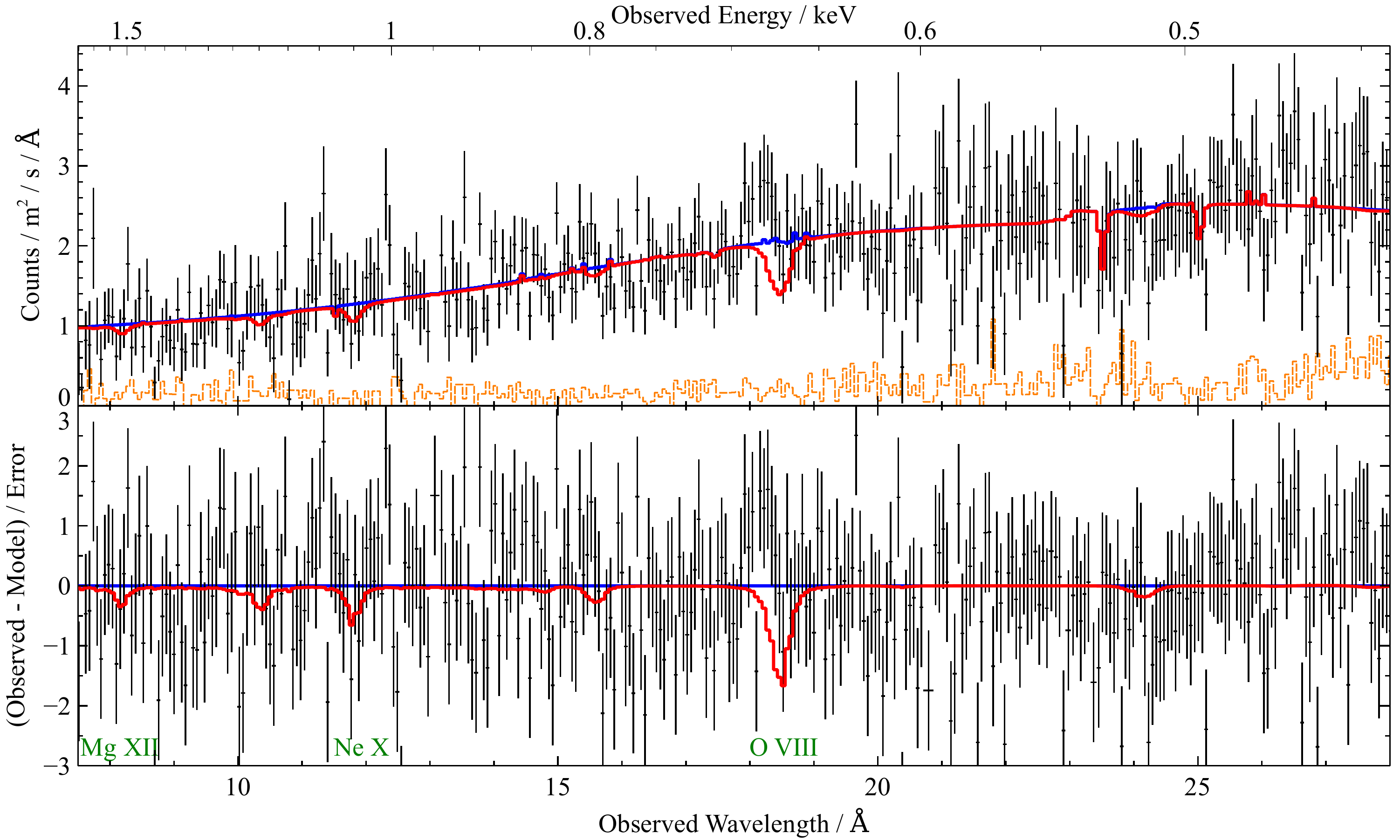}
    \caption{RGS data between 7.5 \AA\ (1.7 keV) and 28 \AA\ (0.4 keV) of the second, short observation of PG 1448+273, stacked and overbinned for plotting purposes only, fitted with the baseline continuum model from Appendix \ref{oldobs} (blue colour). The model in red colour is the best-fitting baseline continuum plus one UFO component with the parameters of the outflow detected in the long \xmm\ observation (Sect. \ref{Sec_xabs_warm}). The top subplot shows the fitted spectra, the bottom subplot contains the residuals. Green labels name the strongest absorption lines of the UFO absorption.}
    \label{RGS_old}
\end{figure*}

Since the statistics are much worse, the addition of neutral absorption intrinsic to the AGN is not significant. The best-fitting spectrum is shown in Fig. \ref{PN_old} (EPIC data) and Fig. \ref{RGS_old} (RGS data). The data are of lower quality than from the other observation, but we do not notice any strong absorption features at 7 keV or at 18 \AA\ as seen in the previous dataset, or any other absorption signature of a UFO. Interestingly, it appears that an emission feature is present at 18 \AA\ instead (Fig. \ref{RGS_old}). If the feature is real, it could be due to \ion{Fe}{XVII} emission (rest-frame wavelength 17.1 \AA) or blueshifted \ion{O}{VIII} emission (rest-frame wavelength 19.0 \AA).

As a further check, we add the same UFO absorption as observed in the long exposure dataset to the baseline continuum of the short observation. We use the best-fitting UFO parameters from Sect. \ref{Sec_xabs_warm}. The resulting spectral model is shown in Fig. \ref{PN_old} and \ref{RGS_old} (red colour), and is clearly rejected by the data. The 1$\sigma$ upper limit on the column density of such absorber is just $1.7 \times 10^{22}$ \pcm, more than 10 times lower than the column density observed in the other \xmm\ observation. We cannot exclude a possibility that a UFO of the same column density is present, but with a higher ionisation parameter. However, the change in the ionisation parameter would have to be significant, with a lower limit on the ionisation parameter of about $\log (\xi / \textrm{erg cm s}^{-1})\approx 5$. Such a large change in the ionisation parameter ($10\times$ higher) would require a non-linear response between the absorber and the ionising luminosity of PG 1448+273.

Finally, we apply the same systematic wind search from Sect. \ref{wind_scan} to locate any potential absorption with parameters different to that of the UFO observed in the other observation. As the data quality is much lower compared to the first exposure and there are no strong spectral residuals, we do not find any significant detections (\delcstat\ $> 15$).

\section{Discussion}
\label{discussion}

We have performed a systematic search for an ionised outflow in the X-ray spectrum of the Narrow Line Seyfert 1 AGN PG 1448+273. We significantly detect a highly ionised wind with a velocity of $(0.090 \pm 0.002)$c.

The absorption lines of the outflow are detected in both hard X-rays (iron K band) and soft X-rays (0.3-2.0 keV, particularly the \ion{O}{VIII} line). This suggests the wind is not as extremely ionised as in other AGN and X-ray binaries where often only \ion{Fe}{XXV/XXVI} features are observed \citep[e.g.][]{Tombesi+10b, Ponti+12}. This fact simplifies the inference of the wind parameters as it reduces the uncertainties in both the velocity and ionisation state of the outflow while simultaneously modelling the wind features as well as the iron K reflection.

The velocity of the outflow of $\sim$0.1c is also not as high as seen in many other AGN \citep[for a review of the recent UFO detections and their velocities see][]{Parker+18}, where velocities as high as 0.3c-0.5c were detected \citep[e.g.][are the most extreme examples]{Reeves+18b, Walton+19}, although velocities in the range of $0.05-0.1c$ have also been detected in some AGN \citep[e.g.][]{Pounds+16a}. Similarly, the velocity width of the observed features is not large at $2100_{-500}^{+600}$ km/s. This suggests that the AGN corona ionising the atoms within the outflow cannot be either too large or too close to where the wind absorption occurs as otherwise the toroidal motion within the wind (imprinted to the wind by the Keplerian rotation of the accretion disc) would broaden the absorption features \citep{Fukumura+19}.

We also note that the outflow velocity is lower than that of the UFO observed in IRAS 13224$-$3809 \citep[0.25c,][]{Parker+16}. At the same time, the inclination of PG 1448+273 ($\sim40^{\circ}$) is lower than the inclination of IRAS 13224$-$3809 \citep[about $70^{\circ}$,][]{Jiang+18}. A lower UFO velocity in PG 1448+273 would be expected in the framework of a model where the UFO absorption is produced in an atmosphere co-rotating with the accretion disc \citep{Fabian+20a} because the projected disc velocities are lower in PG 1448+273.

Curiously, the UFO is not detected in the short (20 ks) 2003 \xmm\ observation of PG 1448+273 (Appendix \ref{oldobs}). The dataset is good enough to significantly reject a UFO of parameters similar to that observed in the main dataset. It is possible that the UFO is still present but at much higher ionisation or with a much lower column density. A change in the outflow velocity alone cannot explain the disappearance of absorption lines from both EPIC and particularly RGS data. Interestingly, PG 1448+273 shows a higher flux during the shorter observation when the UFO is not seen. This is a similar behaviour to that of IRAS 13224$-$3809, which hosts a time-variable UFO, which anti-correlates in absorption strength with the AGN X-ray flux \citep{Parker+16, Pinto+18}. Further long exposures with \xmm\ will be required for a proper flux-resolved analysis to confirm the existence of this important trend in PG 1448+273.

We also detect a weak warm absorber in PG 1448+273. Warm absorbers were detected in multiple AGN with UFO detections, including PG 1211+143 \citep{Reeves+18a}, PG 1114+445 \citep{Serafinelli+19}, IRAS 17020+4544 \citep{Leighly+97, Longinotti+15} and IRAS 13349+2438 \citep{Sako+01, Parker+18b}. Comparing their energetics and mass outflow rates can be helpful in explaining their (possibly common) origin \citep{Tombesi+13} and their interaction with the surrounding environment. Unfortunately, the warm absorber in PG 1448+273 has a very small outflow velocity of $100 \pm 200$ km/s, consistent with being at rest. The limits placed on its energetics are therefore not useful.

We search the X-ray spectrum of PG 1448+273 for further wind structure as multiphase ultra-fast outflows were previously detected in the X-ray spectra of other similar systems such as PG 1211+143 \citep[two outflowing components at 0.06c and 0.13c,][]{Pounds+16a, Pounds+16b, Kriss+18}, 1H 0707$-$495 \citep[0.03c and 0.13c,][]{Kosec+18c} and PDS 456 \citep[$\sim$0.25c and $\sim$0.5c,][]{Boissay-Malaquin+19}. Evidence for even more complex wind structure was found in PG 1114+445 \citep{Serafinelli+19} and in IRAS 17020+4544 \citep{Longinotti+15}. The energetics of the different outflow phases can be used to infer the mode of interaction of the outflow with the surroundings of the AGN. \citet{Kosec+18c} found that three individual outflowing components (including one observed in the UV band) likely have comparable kinetic powers in the NLS1 galaxy 1H 0707$-$495, suggesting energy conservation at these scales.

Our second automated wind search locates a number of tentative outflow phases with a broad range of physical parameters, which would be difficult to identify with manual approach. However, the current data quality prevents us from claiming a significant detection of a multi-phase outflow in the present spectra. Future dedicated campaigns with \xmm\ or with future X-ray observatories, doubling or tripling the total count statistics on PG 1448+273, would be able to constrain these features or rule them out as false detections.

Assuming our measurement of the ionisation parameter of the outflowing material is accurate, we can calculate the kinetic power of the detected fast wind. Following the standard derivations \citep[e.g.][]{Gofford+15}, the mass outflow rate of a disc wind can be expressed as:

\begin{equation}
\dot{M}_{\textrm{out}} = \rho A v = ( C_{\textrm{V}} \mu m_{\textrm{p}} n ) ( \frac{\Omega}{4 \pi} 4 \pi R^{2} ) v
\end{equation}

\noindent where $\rho$ is the gas density and $A$ is the surface area into which the outflow is launched, $v$ is the outflow velocity, $C_{\textrm{V}}$ is the volume filling factor, $\mu$ defines the mean atomic mass ($\sim1.2$ assuming solar abundances), $m_{\textrm{p}}$ the proton mass, $n$ the ion density in the wind. $\frac{\Omega}{4 \pi}$ is the solid angle into which the wind is launched as a fraction of $4 \pi$ and $R$ is the distance of the outflow from the ionising source. The ionising parameter of the wind can be used to get rid of the $R$ and $n$ terms in the equation above since $\xi = \frac{L_{\textrm{ion}}}{nR^2}$, where $L_{\textrm{ion}}$ is the ionising luminosity of the AGN. The mass outflow rate can thus be expressed as:

\begin{equation}
\dot{M}_{\textrm{out}}=C_{\textrm{V}} \Omega \mu m_{\textrm{p}} v \frac{L_{\textrm{ion}}}{\xi}
\end{equation}

\noindent The kinetic power of the outflow is then:

\begin{equation}
\dot{E}_{\textrm{K}}=\frac{1}{2} \dot{M}_{\textrm{out}} v^{2} = \frac{1}{2} C_{\textrm{V}} \Omega \mu m_{\textrm{p}} v^{3} \frac{L_{\textrm{ion}}}{\xi}
\end{equation}

\noindent Using the results from the best-fitting two absorber model in Sect. \ref{Sec_xabs_warm} ($v=26900 \pm 600$ km/s,  $\log (\xi / \textrm{erg cm s}^{-1})=4.03_{-0.08}^{+0.10}$), we find:

\begin{equation}
\dot{E}_{\textrm{K}}= (1.8 \pm 0.5) ~ C_{\textrm{V}} \Omega L_{\textrm{ion}}
\end{equation}

\noindent Even if the clumping factor and the outflow solid angle are relatively small (unlikely as PG 1448+273 is not an edge-on AGN and yet we observe the wind), the kinetic power of the outflow can be comparable to its ionising (13.6 eV $-~13.6$ keV) luminosity. The wind is thus likely more than capable to drive AGN feedback in the host galaxy of PG 1448+273 \citep{diMatteo+05,King+10,Costa+14}. Its energy budget is as large as the UFOs detected in similar NLS1s such as 1H 0707$-$495 \citep{Kosec+18c} and IRAS 13224$-$3809 \citep{Parker+16}.

\section{Conclusions}
\label{conclusions}

Here we summarize our results: 

\begin{itemize}
\item We systematically searched the X-ray spectrum of the Narrow Line Seyfert 1 galaxy PG 1448+273 for signatures of ionised outflows. We unambiguously detect an ultra-fast outflow with a velocity of $26900 \pm 600$ km/s $(0.090^{+0.002}_{-0.002}$c). Our Monte Carlo simulations of the systematic wind search on simulated data show that the UFO detection is highly statistically significant.

\item The wind material has an ionisation parameter of $\log (\xi / \textrm{erg cm s}^{-1})=4.03_{-0.08}^{+0.10}$ and a column density $2.8_{-0.7}^{+1.2} \times 10^{23}$ \pcm. The ionisation of the material is similar to that of outflows found in other AGN. The projected wind velocity is on the lower end of the range  of the extreme outflows observed in other AGN.

\item The spectral features of the ionised outflow are observed in both hard X-rays (iron K band) with EPIC pn and MOS detectors, and in the soft X-ray band with RGS gratings onboard the \xmm\ observatory.

\item Our results underline the importance of the soft X-ray band in detecting high-velocity highly ionised outflows. Inferring the outflow properties from just the iron K band (7-8 keV) is often problematic due to the common presence of the blue wing of the relativistically blurred iron K reflection at similar energies \citep{Zoghbi+15}, as well as due to confusion between the \ion{Fe}{XXV} and \ion{Fe}{XXVI} absorption lines of the ionised outflow.

\item A UFO of similar parameters is not detected in a second, shorter (20ks) observation of PG 1448+273 during which the AGN was in a higher X-ray flux state. The outflow could still be present in the X-ray spectrum, but at a much higher ionisation or with a significantly lower column density. The UFO disappearance at higher fluxes suggests similarities between PG 1448+273 and IRAS 13224$-$3809, where the wind absorption anti-correlates with the X-ray flux \citep{Parker+16}.

\item We also detect a low ionisation warm absorber in soft X-rays. It is consistent with being at rest and has a column density of $(3.8 \pm 1.7) \times 10^{20}$ \pcm\ and an ionisation parameter of $\log (\xi / \textrm{erg cm s}^{-1})=-0.4_{-0.4}^{+0.3}$.

\item We perform a systematic search for further ionisation and velocity phases of the outflow and tentatively detect one or more phases with potential outflow velocities in the range between 20000 and 90000 km/s. However, with the present data quality we are unable to significantly confirm or reject their presence in the spectrum of PG 1448+273.

\end{itemize}

\section*{Acknowledgements}

We are grateful to the anonymous referee for useful comments which improved the clarity and quality of the manuscript. P.K. acknowledges support from the Science and Technology Facilities Council. A.Z. acknowledges support by NASA under Grant No. NNX17AD60G issued through the XMM-Newton GO program. D.J.W. acknowledges support from an STFC Ernest Rutherford Fellowship. C.S.R. thanks the UK Science and Technology Facilities Council for support under the New Applicant grant ST/R000867/1, and the European Research Council for support under the European Union's Horizon 2020 research and innovation programme (grant 834203). This work is based on observations obtained with XMM-Newton, an ESA science mission funded by ESA Member States and USA (NASA). This research has made use of the NASA/IPAC Extragalactic Database (NED), which is funded by the National Aeronautics and Space Administration and operated by the California Institute of Technology.





\bibliographystyle{mnras}
\bibliography{References} 




\appendix

%

\section{Spectral continuum parameters}

\begin{table*}
\def\arraystretch{1.25}
\centering
\caption{Best-fitting X-ray continuum parameters of PG 1448+273 during the main \xmm\ observation. The first column lists the spectral model, the second shows the PG 1448+273 host neutral absorption column density and the third contains the coronal powerlaw slope $\Gamma$. The temperature of the soft excess (described with a blackbody) is in the fourth column. The following 4 columns list the parameters of the blurred iron K line (modelled as a Laor shape): the line energy, the inner disc radius, the emissivity index and the disc inclination, respectively. The final column contains the luminosity in the 2-10 keV band.}
\label{Contparams}
\begin{tabular}{ccccccccc} 
Spectral model&Host N$_{\textrm{H}}$&$\Gamma$&T$_{\textrm{soft}}$&Fe K line energy&R$_{\textrm{in}}$&q&i&L$_{2-10}$\\
&10$^{20}$ \pcm &&keV&keV&R$_{\textrm{G}}$&&&$10^{43}$ erg/s\\

\hline
Continuum only&$9.8 \pm 2.1$&$2.07 \pm 0.03$&$0.097 \pm 0.003$&6.4 (fixed)&$1.52_{-0.29}^{+0.19}$&$4.6 \pm 0.5$&44$_{-3}^{+2}$&$1.40 \pm 0.03$\\
Cont. + UFO + warm. abs.&$6.5^{+2.0}_{-2.2}$&$1.97 \pm 0.03$&$0.102 \pm 0.003$&6.4 (fixed)&$1.7 \pm 0.5$&$3.5^{+2.1}_{-0.5}$&$38^{+16}_{-3}$&$1.81_{-0.10}^{+0.17}$\\

\hline
\end{tabular}
\end{table*}

The best-fitting continuum spectral parameters of PG 1448+273 are listed in Table \ref{Contparams}.

\section{Monte Carlo simulations}
\label{MCSims}

The \delcstat\ value of each point in the systematic multi-parameter wind search grid indicates how strongly the spectral model including the wind absorption is preferred to the baseline continuum model. However, due to the look-elsewhere effect, since we are searching through a large parameter space, it is not trivial to assign directly a `wind detection significance' to a specific value of \delcstat\ \citep{Protassov+02} as their relationship depends on the dataset used and the size of the parameter space. Monte Carlo simulations must be employed to rigorously `map' the C-stat fit improvement to a specific false alarm probability for each wind detection. Such simulations, especially for non-trivial baseline continuum models (as is the case here), can be very expensive if performed over the full parameter space.

Multiple recent wind searches deployed for outflow detection in other objects found with Monte Carlo simulations that $3\sigma$ detection significance corresponded to roughly \delcstat$\sim$20 or $\Delta \chi^2\sim$20 (depending on the fitting method used). The exact values found were $\Delta \chi^2=18.5$ in the analysis of the IRAS 00521$-$7054 AGN \citep{Walton+19}, \delcstat$\sim20$ when studying the ultra-fast outflow of the Ultraluminous X-ray Source (ULX) NGC 1313 X-1 \citep{Pinto+20} and \delcstat$\sim22$ in an Ultraluminous X-ray Source spectrum search \citep{Kosec+18a}. Similarly, \delcstat$\sim30$ fit improvements corresponded to roughly $4\sigma$ significances in ULX wind studies \citep{Kosec+18b, Pinto+20}. Since our UFO detection is above \delcstat\ of 70, the significance of this outflow component is likely very high ($>>3-4\sigma$), comparing with all previous similar analyses.

Nevertheless, we employ Monte Carlo simulations to verify the significance of the UFO detection. The simulations are performed on simulated X-ray spectra with the same exposure and continuum (but no outflow signatures) as the real observation, and should be performed in the same manner as the wind search on real data. Then the fraction of simulated searches with stronger (\delcstat) wind detections than the one in the real data gives the wind detection $p$ value.

However, it is prohibitively expensive to search the whole parameter space many times over. Each MC simulation performed in the same manner as the real wind search (Fig. \ref{2D_Scan_cont}) would take several days on a single (quad-core) computer. We therefore slightly modify the wind search for the MC simulations. We only choose the parameter space between ionisation parameters $\log (\xi/ \textrm{erg cm s}^{-1})=3.5$ and $\log (\xi/ \textrm{erg cm s}^{-1})=4.5$, to cover exactly the properties of the UFO found in the real data (this also covers roughly 1/6 of the full ionisation parameter space). This range of $\log \xi$ is also the region in which most UFOs are detected. Choosing this limited parameter space, we can run a single wind scan for each simulated dataset where the ionisation parameter is allowed to vary in the permitted range, and the turbulent velocity can vary in the 250 km/s to 5000 km/s range, covering most of the velocity width parameter space. The ranges are chosen to be limited so that the routine avoids missing the best-fitting solution at each systematic velocity. The systematic velocity range is the same as in the previous search (20000 km/s to -100000 km/s) with a spacing of 300 km/s. With this set up, each MC simulation takes about 7-8 hours on a single computer.

To test this modified wind search we first apply it on the real data, finding the primary peak with a fit improvement of \delcstat$\sim$83. We note that this is a stronger detection than in the first wind search (where the maximum was \delcstat$\sim73$) as the turbulent velocity of the absorption grid is allowed to vary freely and is not bound to either 100 km/s, 1000 km/s or 10000 km/s. The search is therefore able to find a slightly better fitting solution. 

\begin{figure}
	\includegraphics[width=\columnwidth]{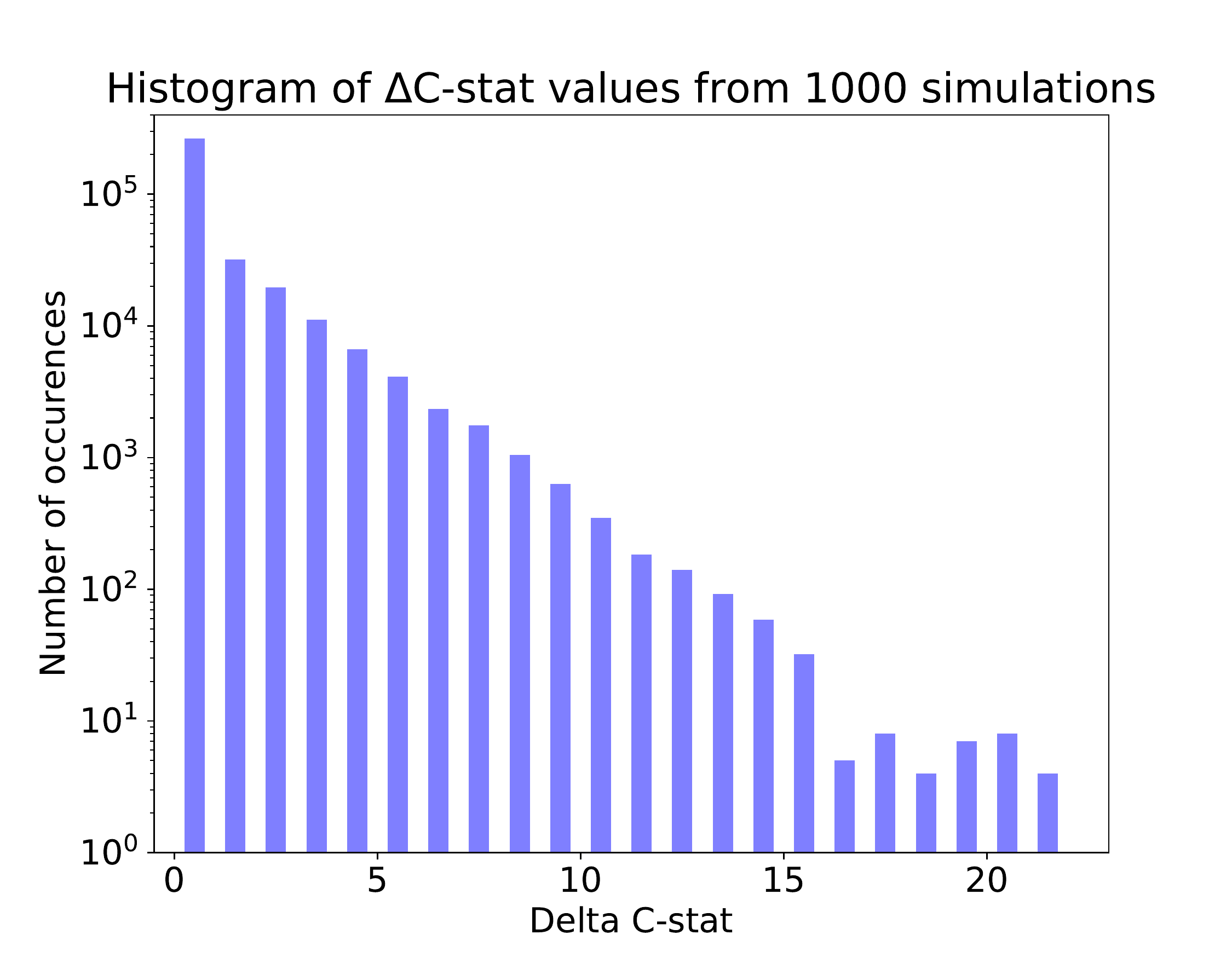}
    \caption{Histogram of \delcstat\ values from 1000 ionised outflow searches on simulated X-ray spectra. All points from the multi-parameter search grids of each simulation are shown.}
    \label{MCsim_hist}
\end{figure}

We performed 1000 Monte Carlo simulations in total, and found no simulated searches with \delcstat>22. The histogram of \delcstat\ values found in the simulated searches is shown in Fig. \ref{MCsim_hist}. The strongest fake detection is \delcstat=21.16, much lower than our wind detection in the real dataset. The $3\sigma$ detection limit with the limited parameter space is at \delcstat$\sim$19.6. Even though the simulated search only covers 1/6 of the full ionisation parameter space, the simulations prove the significance of the UFO detected in the PG 1448+273 spectrum. Assuming the false alarm probability scales linearly with the parameter space (we stress that this is only a very rough assumption), the 95 per cent (2$\sigma$) detection limit is at \delcstat$\sim$16 from our MC simulations.

\bsp	
\label{lastpage}
\end{document}